\documentclass[a4paper, 12pt]{article}
\usepackage[left=2.5cm, right=2.5cm, top=2cm, bottom=2cm]{geometry}
\usepackage{authblk}
\usepackage{graphicx}% Include figure files
\usepackage{adjustbox}
\usepackage{amsmath}
\usepackage{dcolumn}% Align table columns on decimal point
\usepackage{bm}% bold math
\usepackage{hyperref}
\usepackage{xspace}                     
\usepackage{comment}
\usepackage{multirow}
\usepackage{cite}
\usepackage{xcolor}
\usepackage{comment}

\begin{document}

\title{A toy model for understanding the space-point resolution of silicon pixel detectors with digital readout}
\author[1]{Marianna Mazzilli}
\author[2]{Alexander Kalweit}

\affil[1]{\small Dipartimento Interateneo di Fisica M. Merlin and Sezione INFN, Bari, Italy}
\affil[2]{Experimental Physics department, CERN}
\date{}

%\linenumbers

\maketitle

\begin{abstract}
Silicon pixel detectors are widely used in high-energy physics experiments as tracking detectors close to the primary interaction vertex. They provide excellent space-point resolution together with fast electronic readout. Many of them employ only digital (binary) readout which makes the applicability of centre-of-gravity algorithms less obvious. Charge sharing between neighboring pixels improves the resolution beyond the single-pixel limit, but in practice there is a lack of quantitative understanding of the achievable gains. This work provides a simplified analytical and numerical model with which the maximum improvement achievable through charge sharing is quantified for both one and two-dimensional pixel geometries. A phenomenological parameterisation of the resolution as a function of the average cluster size is derived and compared to experimental data from several detector technologies.

\end{abstract}

\section{Introduction}

One of the key performance parameters of modern tracking detectors is the space-point resolution $\sigma_p$. In particular, it directly enters into the momentum and distance-of-closest approach (DCA) resolution of charged particles. The DCA resolution $d_0$ is the decisive quantity in identifying charm and beauty hadron decays. At high momenta, where multiple scattering is negligible, it is given by~\cite{Musa:2025yhp}:

\begin{equation}
 d_0 \approx \frac{3\sigma_p}{\sqrt{N+4}} \; ,
\end{equation}

\noindent where $N$ denotes the number of layers of the tracking detector. The optimization of $\sigma_p$ is therefore of paramount importance in the detector and chip design. For detectors with analogue readout, the hit position is typically reconstructed as the weighted average of pixel positions, using the signal amplitudes as weights. In this case, it is immediately obvious that the space-point resolution $\sigma_p$ improves with increasing cluster size $N_{clus}$, as progressively more information is incorporated. The dependence of spatial resolution on charge sharing is in this case largely textbook knowledge (see for instance~\cite{Rossi:2006buh}). 

%% TO-DO-HERE: referee comment #1 -> add discussion and references that for analogue readout, this is all known and well studied

For a detector with digital (binary) readout, the situation is less obvious since all pixels carry equal weight. As will be seen later, in this case the space point resolution can also worsen with increasing cluster size beyond a certain minimum. Such effects are to our knowledge not well understood and are not discussed in the same level of detail as in the present work.

A detector with binary readout is, for instance, realised in the ALPIDE chip employed in the Inner Tracking System~2 (ITS2) of the ALICE experiment at the LHC~\cite{Mager:2016yvj}. In this paper, we develop a simplified model -- both analytically and via toy Monte Carlo simulations -- that identifies the average cluster size as a scaling variable, quantifies the ideal (maximal) resolution improvement achievable through charge sharing in detectors with binary readout, and compares to recent data from existing silicon pixel detectors. 

Related studies exist, but they differ in focus and mostly do not compare their results with experimental data. Wang et al.~\cite{Wang:2017oip} convolve the achievable resolution of several fully simulated, fixed detector geometries with the information on the particle’s incident angle, as provided by multi-layer tracking detectors. Yonamine et al.~\cite{Yonamine:2016aoa} focus on one-dimensional geometries and explicitly account for drift regions. Turchetta~\cite{Turchetta:1993vu} compares with experimental data, but only at a fixed threshold and as a function of the incident angle. Our study, in contrast, addresses perpendicular incidence with a varying threshold.

\section{Analytic modeling}

\subsection{Single-pixel clusters}

\begin{figure}[h]
 \begin{center}

   \includegraphics[width=\linewidth]{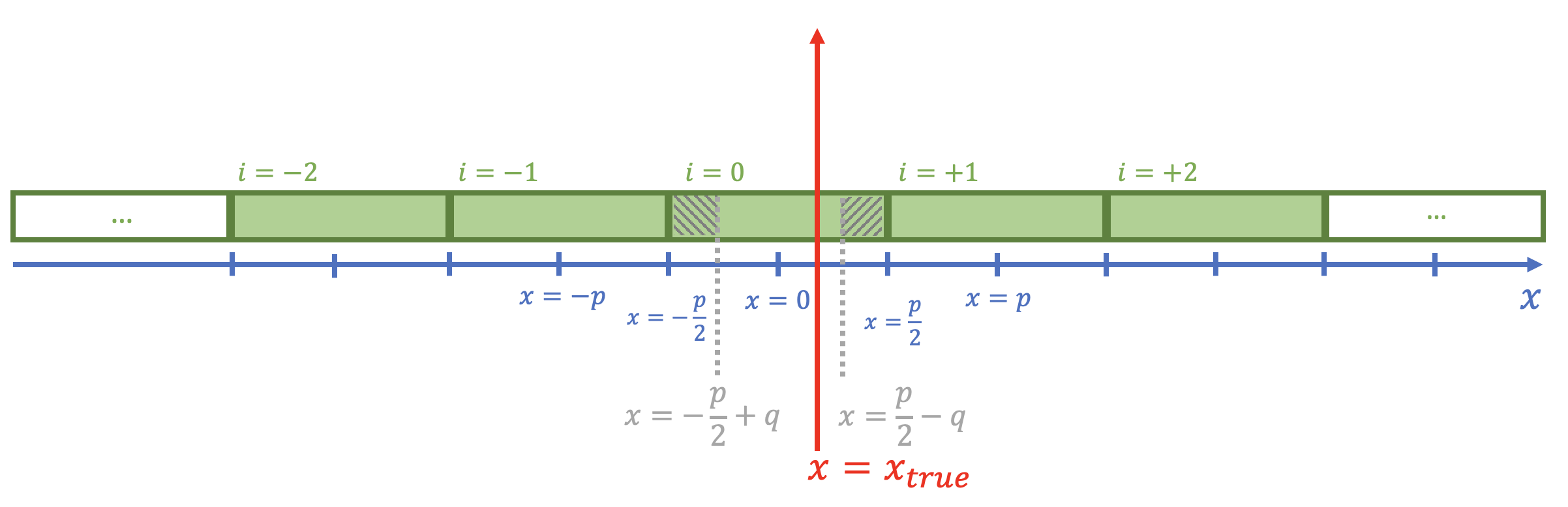}

  \caption{Sketch of the one-dimensional geometry. The pixel matrix with pitch $p$ is indicated by the green boxes enumerated by the index $i$. A charged particle (red arrow) crosses the detector at position $x_{true}$. }
  \label{Fig1Danalytic}
 \end{center}
\end{figure}

For many silicon pixel detectors, the limiting case of single-pixel (one-pixel) clusters is often used as a reference. We review it here to set the stage for the following derivations. We start with the one-dimensional case in which a detector along the $x$-direction is segmented into pixels with a uniform pitch $p$ as shown in Fig.~\ref{Fig1Danalytic}.

For simplicity, we consider here and throughout the paper only trajectories corresponding to particles impinging perpendicular on the detector plane at a position $x_{true}$. In addition, we assume in this chapter that the number of fired pixels $N_{pixels}$ in a given event does depend solely on the impact point of the impinging particle $x_{true}$. We thus do not explicitly describe the distribution of the created charge itself. This will only be done in the following chapter where a Monte-Carlo toy model is employed.

By symmetry, all cases can be described by considering only particles hitting the central pad ($i=0$), where $i$ is the pad index. The probability density $f(x_{true})$ of the impinging position is described by a uniform distribution:

\begin{equation}
  f(x_{true}) = \begin{cases} 1/p & -p/2 < x_\text{true} < p/2 \\ 0 & otherwise \end{cases} \;\; . \label{Eq.:UniformParticleDistribution}
\end{equation}

\noindent The reconstructed position $x_{rec}$ is given by the weighted average of the pad centres $x_{i} = ip$, using the signal $S(i,x_{true})$ on pad $i$ for a given $x_{true}$ as weight:

\begin{equation}
  x_\text{rec} = \frac{\sum_{i}S(i,x_{true}) \cdot ip}{\sum_{i}S(i,x_{true})} \; .
\end{equation}

\noindent For a one-pixel cluster, we consider a strictly point-like energy deposition at $x_{true}$ that leads to a signal that is defined as

\begin{equation}
  S(i, x_{true}) = \begin{cases} 1 & -p/2 < x_\text{true} < p/2, i = 0 \\ 0 & otherwise \end{cases} \;\; .
\end{equation}

\noindent The reconstructed position therefore reduced to the centre of the central pixel $x_\text{rec} = 0$. For the resolution $\sigma_p$, we obtain

\begin{eqnarray}
\sigma_p^2 &=&\int_{-p/2}^{p/2} (x_\text{rec} - x_{true})^2 f(x_{true})~{\mathrm d}x_\text{true} = \int_{-p/2}^{p/2} x_{true}^2 \frac{1}{p} ~{\mathrm d}x_\text{true} \\
&=& \frac{1}{p} \left [\frac{1}{3} x_{true}^3\right ]_{-p/2}^{p/2} = \frac{1}{3p} \left(\frac{p^3}{8} + \frac{p^3}{8}\right) \\
&=& \frac{p^2}{12} \; .
\end{eqnarray}

\noindent This yields the well-known result $\sigma_p = p/\sqrt{12}$ relation for one-pixel clusters.

\subsection{One- and two-pixel clusters}

We now discuss the specific case in which both one and two pixel clusters occur. In this scenario, we assume that if a track passes through a given pixel $i$, only this pixel fires if the track is close to the centre of this pixel. If it is close to the boundary of the next pixel, e.g. $i+1$, both pixels fire. As illustrated in Fig.~\ref{Fig1Danalytic}, this behaviour is characterised by a specific minimal distance $q$ from the pixel boundary beyond which charge sharing occurs. Analogously to the one pixel case, we start identical as in Eq.~\ref{Eq.:UniformParticleDistribution} with a uniform distribution:

\begin{equation}
  f(x_{true}) = \begin{cases} 1/p & -p/2 < x_\text{true} < p/2 \\ 0 & otherwise \end{cases} \;\; .
\end{equation}

\noindent The reconstructed position depends on whether the neighboring pixels fire. Specifically: 

\begin{itemize}
\item For $-p/2 < x_\text{true} < -p/2 +q$, the pixels $i=-1$ and $i=0$ fire. 
\item For $-p/2 + q < x_\text{true} < p/2 -q$ only the central pixel $i=0$ fires. 
\item For $p/2 - q < x_\text{true} < p/2$ the pixels $i=0$ and $i=1$ fire. 
\end{itemize}
We therefore obtain for the weighted average

\begin{equation}
  x_\text{rec} = \begin{cases} -p/2 & -p/2 < x_\text{true} < -p/2+q \\ 0 & -p/2+q < x_\text{true} < p/2-q \\ p/2 &~~ p/2-q < x_\text{true} < p/2 \end{cases} \;\; .
\end{equation}

\noindent For the corresponding resolution, we then obtain

\begin{eqnarray}
\sigma_p^2 &=&\int_{-p/2}^{p/2} (x_\text{true} - x_{rec})^2 f(x_{true}) ~{\mathrm d}x_\text{true} \\ &=&\int_{-p/2}^{-p/2+q} (x_{true}+p/2)^2 \frac{1}{p}~ {\mathrm d}x_\text{true} +\\ &+& \int_{-p/2+q}^{p/2-q} x_{true}^2 \frac{1}{p}~ {\mathrm d}x_\text{true} + \int_{p/2-q}^{p/2} (x_{true}- p/2)^2 \frac{1}{p}~ {\mathrm d}x_\text{true} 
\end{eqnarray}

\noindent with the substitutions $y= x + p/2$ and $z= x-p/2$ this reduces to

\begin{eqnarray}
\sigma_p^2 &=& \int_{0}^{q} y^2 \frac{1}{p}~ {\mathrm d}y + \int_{-p/2+q}^{p/2-q} x_{true}^2 \frac{1}{p}~ {\mathrm d}x_\text{true} + \int_{-q}^{0} z^2 \frac{1}{p}~ {\mathrm d}z \\
&=& {1 \over p} \left [{1 \over 3} y^3\right ]_{0}^{q}+{1 \over p} \left [{1 \over 3} x_{true}^3\right ]_{-p/2+q}^{p/2-q} + {1 \over p} \left [{1 \over 3} z^3\right ]_{-q}^{0} \\
&=& \frac{p^2 - 6pq + 12q^2}{12} \; \label{Eq.:Parabola}.
\end{eqnarray}

\noindent Setting $\mathrm{d}\sigma_p^2/\mathrm{d}q = 0$, we find that the optimal (minimal) resolution is achieved at $q=p/4$, yielding

\begin{equation}
    \sigma_p = \frac{1}{2\sqrt{12}}p \approx 0.144p \; ,
    \label{Eq.:bestResolution}
\end{equation}

\noindent i.e. exactly half of the one-pixel case. This result has an intuitive interpretation: if one knows that only one pixel fires for $-p/4 < x_\text{true} < p/4$ and two fire for $p/4 < x_\text{true} < p/2$, the particle position is effectively constrained to half the pixel pitch. The non-trivial aspect of this result is that it represents the best achievable resolution. As we will see in the following, this limit also  holds for larger cluster sizes in the one-dimensional case.

The case $q=p/4$ corresponds in practice to an average cluster size $\langle N_\text{pixels} \rangle$ of $\langle N_\text{pixels} \rangle=1.5$, since in half of the cases one pixel fires and in the other half two pixels fire. As a matter of fact, there is a one-to-one correspondence between $q$ and the average cluster size that is given by $\langle N_\text{pixels} \rangle = 1 +\frac{2q}{p}$, i.e. the ratio between $q$ and the pad size. We will formalise this relation in the next chapter in order to directly express the resolution as a function of the average cluster size.

\subsection{Arbitrary cluster sizes}

For the investigation of larger cluster sizes, we generalize the previous case under the assumption that the charge is shared symmetrically around $x_\text{true}$. In this situation, it becomes obvious that for a large enough charge, for $-p/2 + q < x_\text{true} < p/2 -q$ the central pixel $i=0$ and its two neighboring pixels $i=-1$ and $i=+1$ fire with consequently $x_\text{rec} = 0$. Moving $x_\text{true}$ to larger values of $x$ will lead to a decrease of charge in $i=-1$ so that for $p/2 - q < x_\text{true} < p/2$ the pixels $i=0$ and $i=1$ fire with consequently $x_\text{rec} = p/2$. Likewise, we get for  $- p/2  < x_\text{true} < -p/2 + q$ that $x_\text{rec} = -p/2$. In other words, we obtain the same behavior as in the previous case.

To facilitate comparison with the toy model results presented in the next section and to allow processing with computer algebra programs, we express this formally as a function of the average cluster size $\langle N_\text{pixels} \rangle$ with the help of the Heaviside step function $\Theta$. The uniform probability density of $x_\text{true}$ over the central pad is written as

\begin{equation}
f(x_{\text{true}}) = \frac{1}{p} \left( \Theta\!\left(x_{\text{true}} + \frac{p}{2}\right) - \Theta\!\left(x_{\text{true}} - \frac{p}{2}\right) \right) \; .
\end{equation}

\noindent As there is a one-to-one correspondence between the average cluster size $\langle N_\text{pixels} \rangle$ and the previously employed variable $q$, we directly describe the signal $S(i)$ (which is either 0 or 1) on pad $i$, whose centre is at $x_{pc} = i \cdot p$ as

\begin{equation}
S(i,\, x_{\text{true}},\, \langle N_\text{pixels} \rangle) = \Theta\!\left(x_{\text{true}} - \left(x_{\text{pc}}(i) - \frac{\langle N_\text{pixels} \rangle \cdot p}{2}\right)\right) - \Theta\!\left(x_{\text{true}} - \left(x_{\text{pc}}(i) + \frac{\langle N_\text{pixels} \rangle \cdot p}{2}\right)\right) \;.
\end{equation}

\noindent One can then indeed verify that the average cluster size is given by

\begin{equation}
\langle N_\text{pixels} \rangle = \frac{1}{p} \int_{-p/2}^{p/2} \sum_{i=-\infty}^{+\infty} S(i,\, x_{\text{true}},\, \langle N_\text{pixels} \rangle) \, dx_{\text{true}} \; .
\end{equation}

\noindent Based on the reconstructed position
\begin{equation}
x_{\text{rec}}(x_{\text{true}},\, \langle N_\text{pixels} \rangle) = \frac{\displaystyle\sum_{i=-\infty}^{+\infty} x_{\text{pc}}(i) \cdot S(i,\, x_{\text{true}},\, \langle N_\text{pixels} \rangle)}{\displaystyle\sum_{i=-\infty}^{+\infty} S(i,\, x_{\text{true}},\, \langle N_\text{pixels} \rangle)}
\end{equation}

\noindent we then obtain for the resolution

\begin{eqnarray}
 \sigma_p^2 &=& \int_{-p/2}^{p/2} \Bigl( x_{\text{rec}}(x_{\text{true}},\, \langle N_\text{pixels} \rangle) - x_{\text{true}} \Bigr)^2 \cdot f(x_{\text{true}}) \, dx_{\text{true}} \nonumber \\
 &=& \begin{cases}
\dfrac{p^2}{12} & \langle N_\text{pixels} \rangle = 1,\, 2,\, 3,\, 4,\, 5,\, 6,\, 7,\, 8,\, 9,... \\[10pt]
\dfrac{1}{12}\left(7 - 9\,\langle N_\text{pixels} \rangle + 3\,\langle N_\text{pixels} \rangle^2\right) p^2    & 1 < \langle N_\text{pixels} \rangle < 2 \\[10pt]
\dfrac{1}{12}\left(19 - 15\,\langle N_\text{pixels} \rangle + 3\,\langle N_\text{pixels} \rangle^2\right) p^2  & 2 < \langle N_\text{pixels} \rangle < 3 \\[10pt]
\dfrac{1}{12}\left(37 - 21\,\langle N_\text{pixels} \rangle + 3\,\langle N_\text{pixels} \rangle^2\right) p^2  & 3 < \langle N_\text{pixels} \rangle < 4 \\[10pt]
\dfrac{1}{12}\left(61 - 27\,\langle N_\text{pixels} \rangle + 3\,\langle N_\text{pixels} \rangle^2\right) p^2  & 4 < \langle N_\text{pixels} \rangle < 5 \\[10pt]
\dfrac{1}{12}\left(91 - 33\,\langle N_\text{pixels} \rangle + 3\,\langle N_\text{pixels} \rangle^2\right) p^2  & 5 < \langle N_\text{pixels} \rangle < 6 \\[10pt]
\dfrac{1}{12}\left(127 - 39\,\langle N_\text{pixels} \rangle + 3\,\langle N_\text{pixels} \rangle^2\right) p^2 & 6 < \langle N_\text{pixels} \rangle < 7 \\[10pt]
...
\end{cases} \; \label{Eq.:SigmaVsClusSizeAnalytic}.
\end{eqnarray}

\noindent While this piecewise expression appears complicated at first glance, it is simply the quadratic polynomial from Eq.~\ref{Eq.:Parabola} translated piecewise and expressed as a function of the average cluster size. Each parabolic segment has its minium at $\sigma_p = \frac{1}{2\sqrt{12}}p$, confirming that the two-pixel limit from Eq.~\ref{Eq.:bestResolution} is the best achievable resolution regardless of the cluster size in one dimension. In the following section, we compare this analytical expression to toy model results.

\section{Toy models of pixel detectors}

To investigate the interplay between charge sharing, clustering, threshold effects, and spatial resolution under binary readout, two simplified detector models were implemented: a one-dimensional pixel array and a fully segmented two-dimensional pixel matrix. Both models are noise-free and isolate the geometric and threshold-driven effects on cluster formation and position reconstruction.

\begin{figure}[htbp]
\centering

\includegraphics[width=0.65\textwidth]{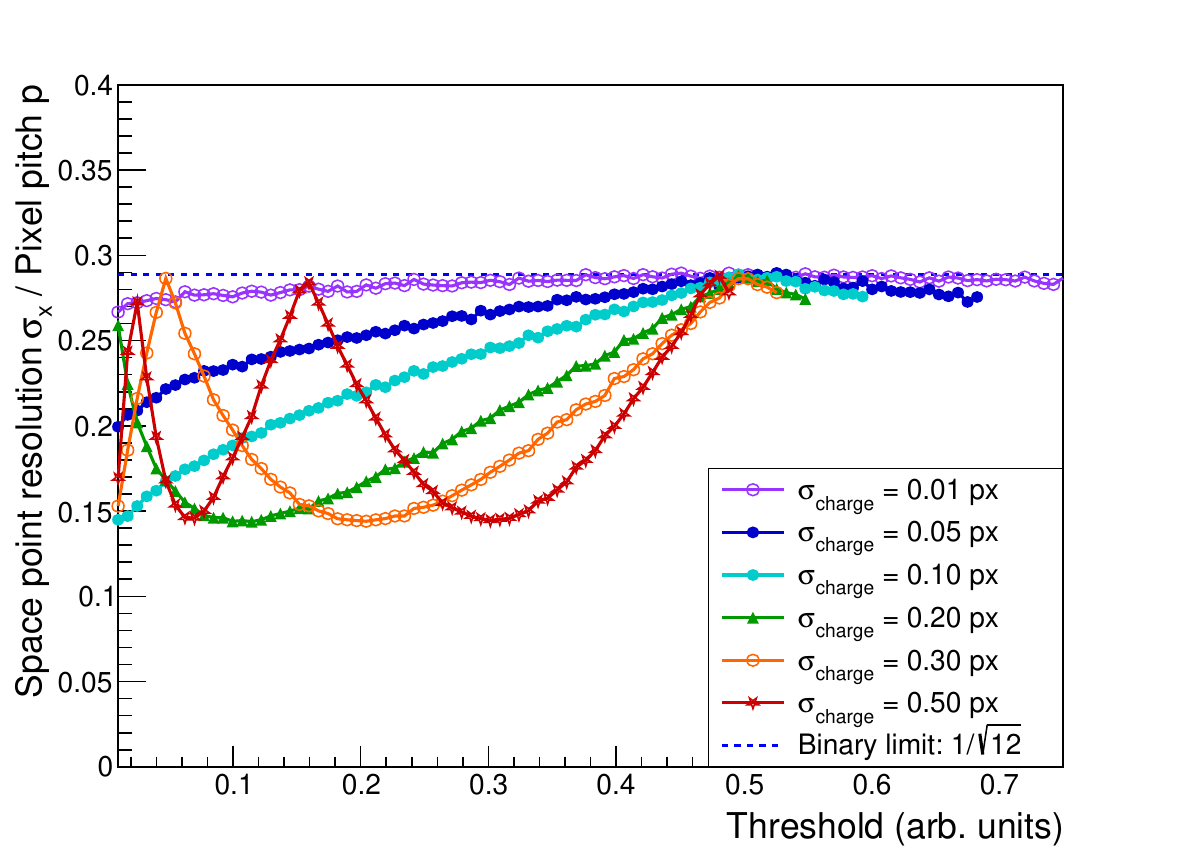}
\includegraphics[width=0.65\textwidth]{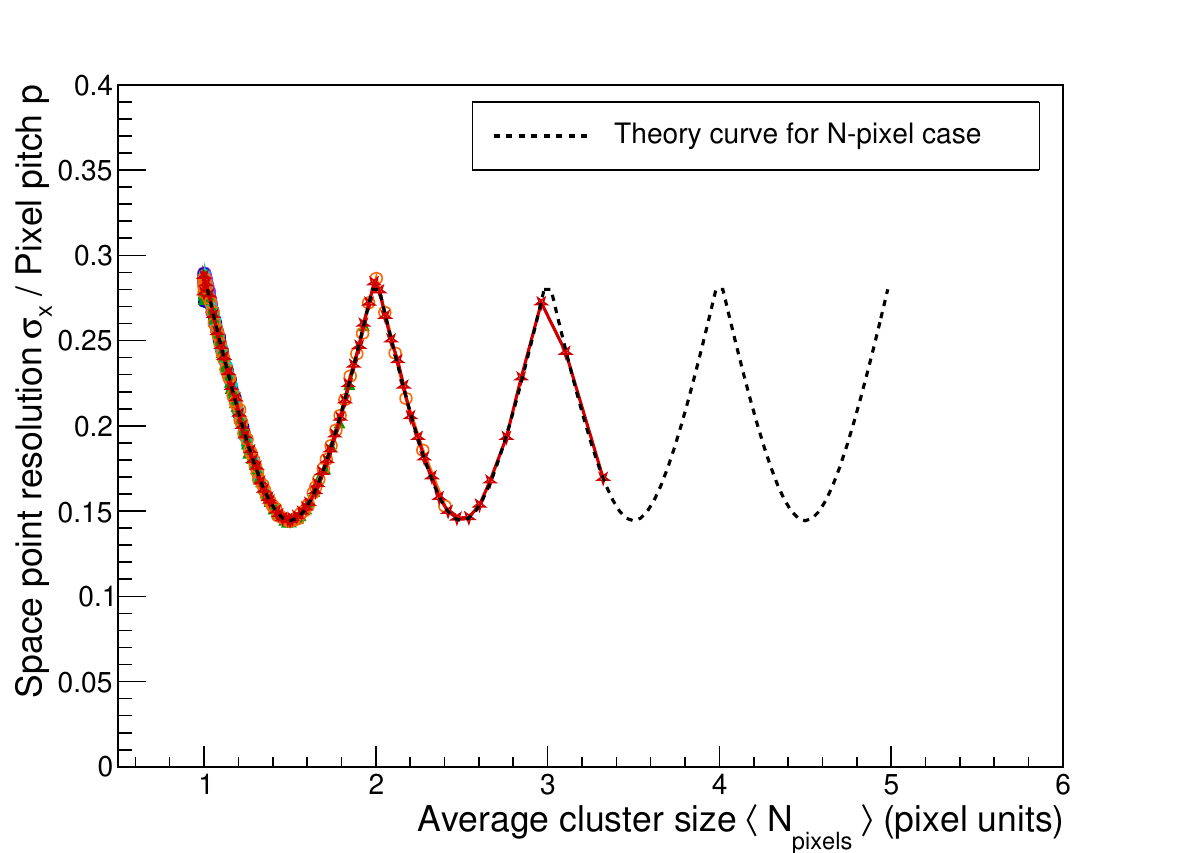}
\includegraphics[width=0.65\textwidth]{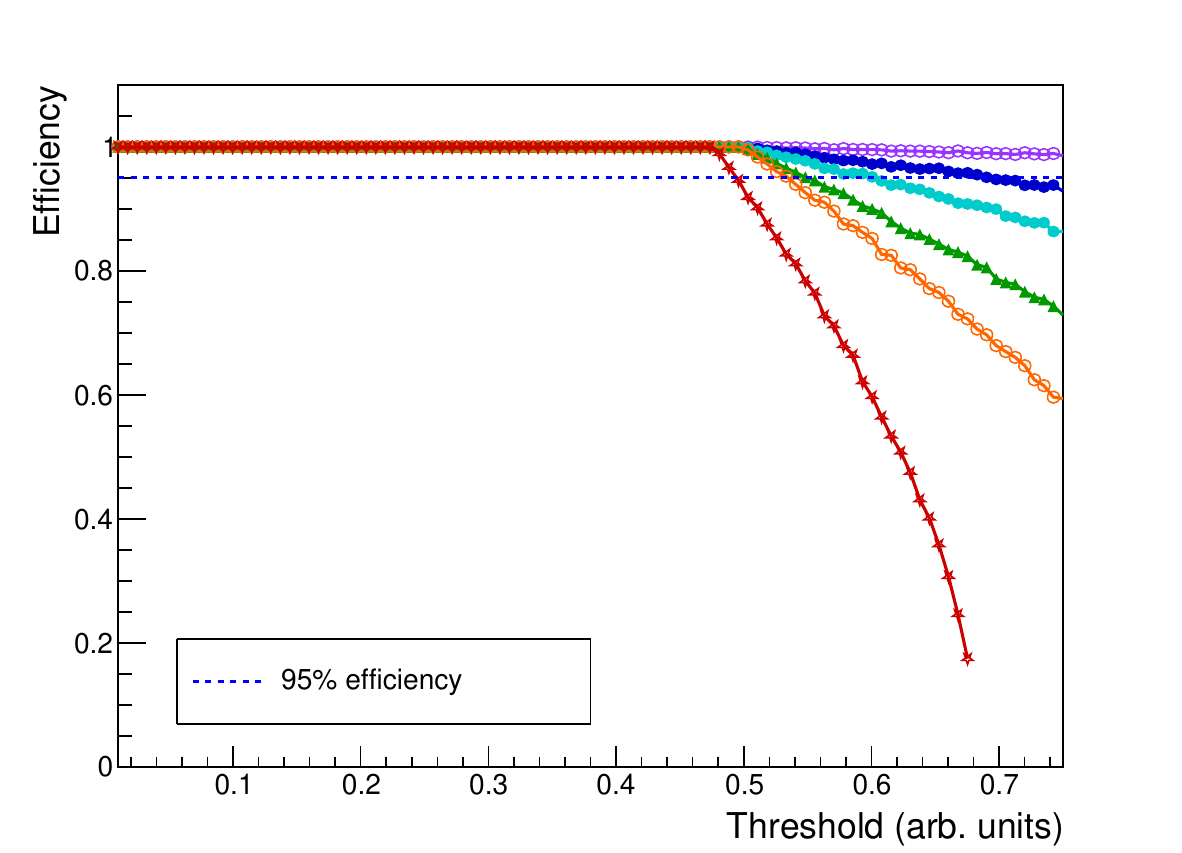}
\caption{Performance of the one-dimensional toy model. Top: spatial resolution as a function of the pixel threshold. Middle: the same spatial resolution values, but plotted as a function of the average cluster size. Results are compared to the theoretical curve from Eq.~\ref{Eq.:SigmaVsClusSizeAnalytic}. Bottom: detection efficiency versus threshold. Resolution plots are restricted to threshold values where the efficiency exceeds 95\%.}
\label{fig:1D_performance}
\end{figure}

\subsection{One-Dimensional Pixel Model}

The one-dimensional model consists of a linear array of $n$ pixels with uniform pitch $p = 1$ (in pixel-pitch units), centred at $x = 0$. For practical purposes, we chose $n=21$ in order to ensure $n \gg 1$. A particle crossing the detector at position $x_{\mathrm{true}}$ deposits charge according to a Gaussian distribution with transverse width $\sigma_{\mathrm{charge}}$, expressed in units of pixel pitch $p$. The charge collected by each pixel is computed by analytically integrating the Gaussian over the pixel boundaries,
\begin{equation}
Q_i =
\frac{1}{2}
\left[
\operatorname{erf}\!\left(
\frac{x_i - x_{\mathrm{true}} + p/2}{\sqrt{2}\sigma_{\mathrm{charge}}}
\right)
-
\operatorname{erf}\!\left(
\frac{x_i - x_{\mathrm{true}} - p/2}{\sqrt{2}\sigma_{\mathrm{charge}}}
\right)
\right],
\end{equation}
which ensures proper normalization and realistic charge sharing between neighboring pixels. No electronic noise is included. This normalization implies that the total deposited charge, summed over all pixels, is unity ($Q_{\text{tot}} = 1$, in arbitrary charge units). The threshold $Q_{\mathrm{thr}}$ is therefore expressed in the same units and can be directly interpreted as the fraction of the total charge that a pixel must collect in order to be marked as active.

A fixed per-pixel threshold $Q_{\mathrm{thr}}$ is applied independently to each channel, and pixels with $Q_i > Q_{\mathrm{thr}}$ are marked as active. Clusters are formed by identifying the first pixel above threshold and grouping all contiguous pixels above threshold around it. Only one cluster per event is reconstructed.

The reconstructed position is determined using a binary centroid algorithm,
\begin{equation}
x_{\mathrm{rec}} = \frac{1}{N_\mathrm{pixels}} \sum_{i \in \mathrm{cluster}} x_i,
\end{equation}
where $N_\mathrm{pixels}$ is the cluster size. The spatial resolution is defined as the root-mean-square of the residual distribution,
\begin{equation}
\sigma_x = \sqrt{\left\langle (x_{\mathrm{rec}} - x_{\mathrm{true}})^2 \right\rangle}.
\end{equation}
Events are generated with $x_{\mathrm{true}}$ uniformly distributed within the central pixel. This minimal geometry reproduces the expected binary limit $\sigma_x = 1/\sqrt{12}$ for single-pixel clusters and allows to study the spatial resolution for simplified cluster topologies. 

The performance of the model is studied by scanning the pixel threshold and evaluating the resulting cluster properties. The dependence of the spatial resolution on the threshold, together with the average cluster size and detection efficiency, provides insight into the mechanisms governing position reconstruction with binary readout. The detection efficiency is defined as the fraction of simulated events for which at least one pixel collects a charge above the discrimination threshold, thereby forming at least one cluster, i.e., $\varepsilon = N_{\text{reco}}/N_{\text{gen}}$, where $N_{\text{gen}}$ is the total number of generated events and $N_{\text{reco}}$ is the number of events with at least one reconstructed cluster. Figure~\ref{fig:1D_performance} summarizes these quantities, showing the resolution as a function of threshold for efficiencies larger than 95\%, the resolution as a function of the average cluster size for efficiencies larger than 95\%, and the detection efficiency versus threshold. 

As already indicated by the analytic calculations, the best possible resolution amounts to $\frac{p}{2\sqrt{12}}$. This limit is essentially respected for all cluster sizes. Similarly, the worst resolution corresponds to the one pixel limit $\frac{p}{\sqrt{12}}$.  We also show the comparison to the analytic expression of Eq.~\ref{Eq.:SigmaVsClusSizeAnalytic} in Fig.~\ref{fig:1D_performance} and an excellent agreement is found.

\subsection{Two-dimensional pixel model}

The two-dimensional model extends the geometry to a $21 \times 21$ matrix of square pixels with pitch $p = 1$, centred at $(x,y) = (0,0)$. A particle crossing the detector at $(x_{\mathrm{true}}, y_{\mathrm{true}})$ deposits charge according to a two-dimensional Gaussian with width $\sigma_{\mathrm{charge}}$. The collected charge per pixel is computed by integrating the Gaussian over the pixel area. Owing to the separability of the Gaussian, the integrated charge factorizes as
\begin{equation}
Q_{ij} = Q_x(i)\,Q_y(j),
\end{equation}
where $Q_x$ and $Q_y$ have the same error-function form as in the one-dimensional case. This approach provides an analytically exact description of charge sharing in both directions.

%A per-pixel threshold $Q_{\mathrm{thr}}$ is applied, and clusters are identified using a two-dimensional connected-component algorithm based on four-neighbor connectivity. All contiguous pixels above threshold are grouped into clusters. 

A per-pixel threshold $Q_{\mathrm{thr}}$ is applied independently to each pixel, and clusters are identified using a two-dimensional connected-component algorithm based on four-neighbor connectivity, i.e., only pixels sharing an edge (and not diagonally adjacent pixels) are considered part of the same cluster. All contiguous pixels above threshold are grouped into clusters; when multiple clusters are found in an event, only the one with the largest total charge is retained for position reconstruction. The cluster size is defined as the total number of pixels above threshold belonging to this cluster.

The reconstructed position is obtained from a binary centroid in both coordinates,
\begin{equation}
x_{\mathrm{rec}} = \frac{1}{N_\mathrm{pixels}} \sum_{k \in \mathrm{cluster}} x_k,
\qquad
y_{\mathrm{rec}} = \frac{1}{N_\mathrm{pixels}} \sum_{k \in \mathrm{cluster}} y_k,
\end{equation}
where $N_\mathrm{pixels}$ is the total number of pixels in the cluster. The spatial resolutions are defined as
\begin{equation}
\sigma_x = \sqrt{\left\langle (x_{\mathrm{rec}} - x_{\mathrm{true}})^2 \right\rangle},
\qquad
\sigma_y = \sqrt{\left\langle (y_{\mathrm{rec}} - y_{\mathrm{true}})^2 \right\rangle}.
\end{equation}

Events are generated with $(x_{\mathrm{true}}, y_{\mathrm{true}})$ uniformly distributed within the central pixel. Compared to the one-dimensional case, the two-dimensional geometry naturally produces larger clusters due to charge sharing in both coordinates and allows a realistic study of cluster topology, efficiency, and resolution under binary readout. Together, the 1D and 2D models provide complementary frameworks for isolating and understanding the impact of thresholding and charge diffusion on position reconstruction performance in segmented pixel detectors.\\

\begin{figure}[t]
\centering
\includegraphics[width=0.495\textwidth]{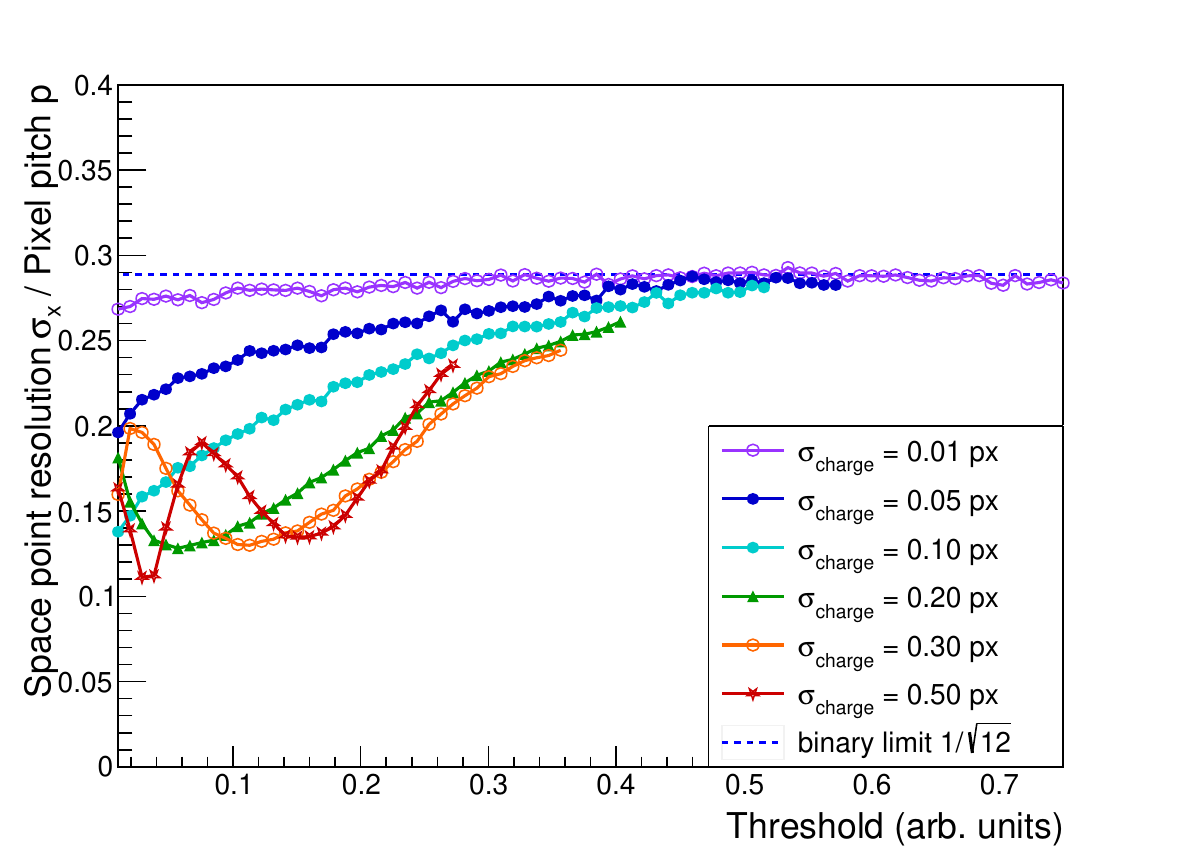}
\includegraphics[width=0.495\textwidth]{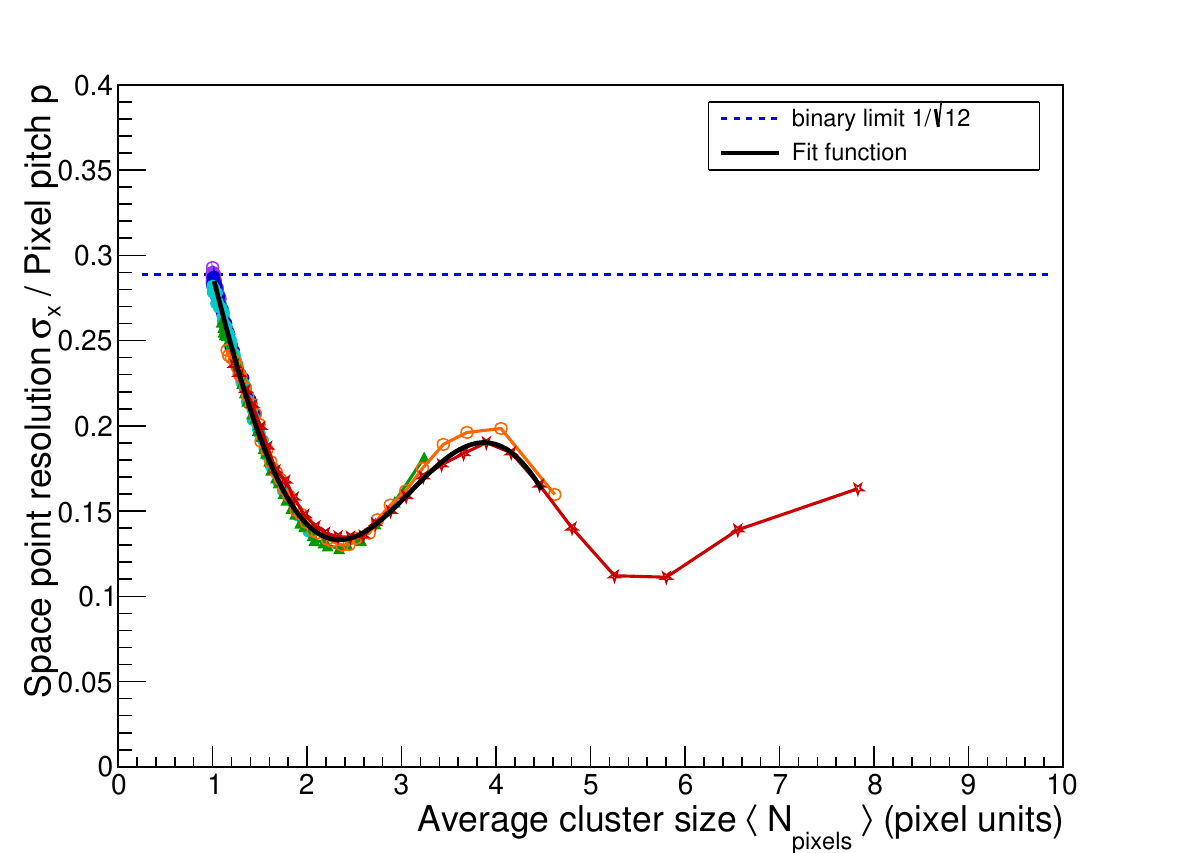}

\vspace{0.5em}

\includegraphics[width=0.495\textwidth]{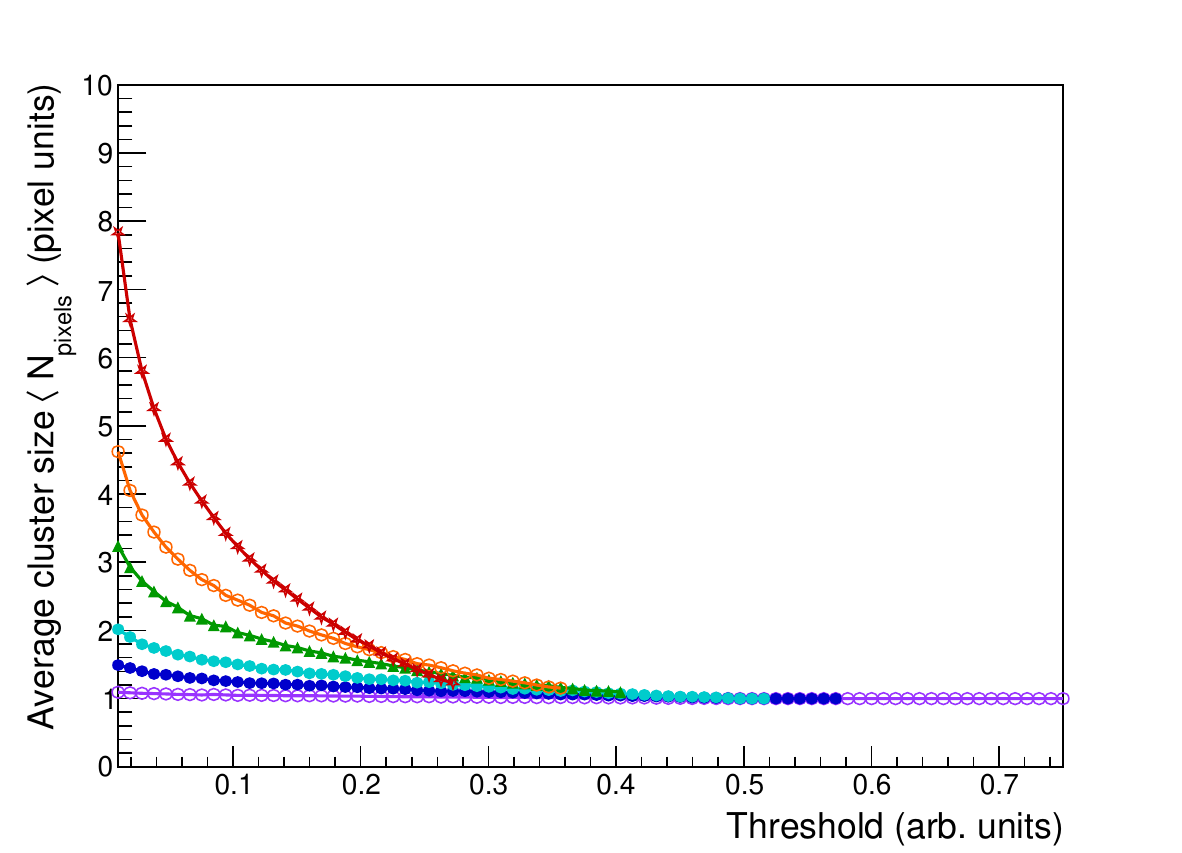}
\includegraphics[width=0.495\textwidth]{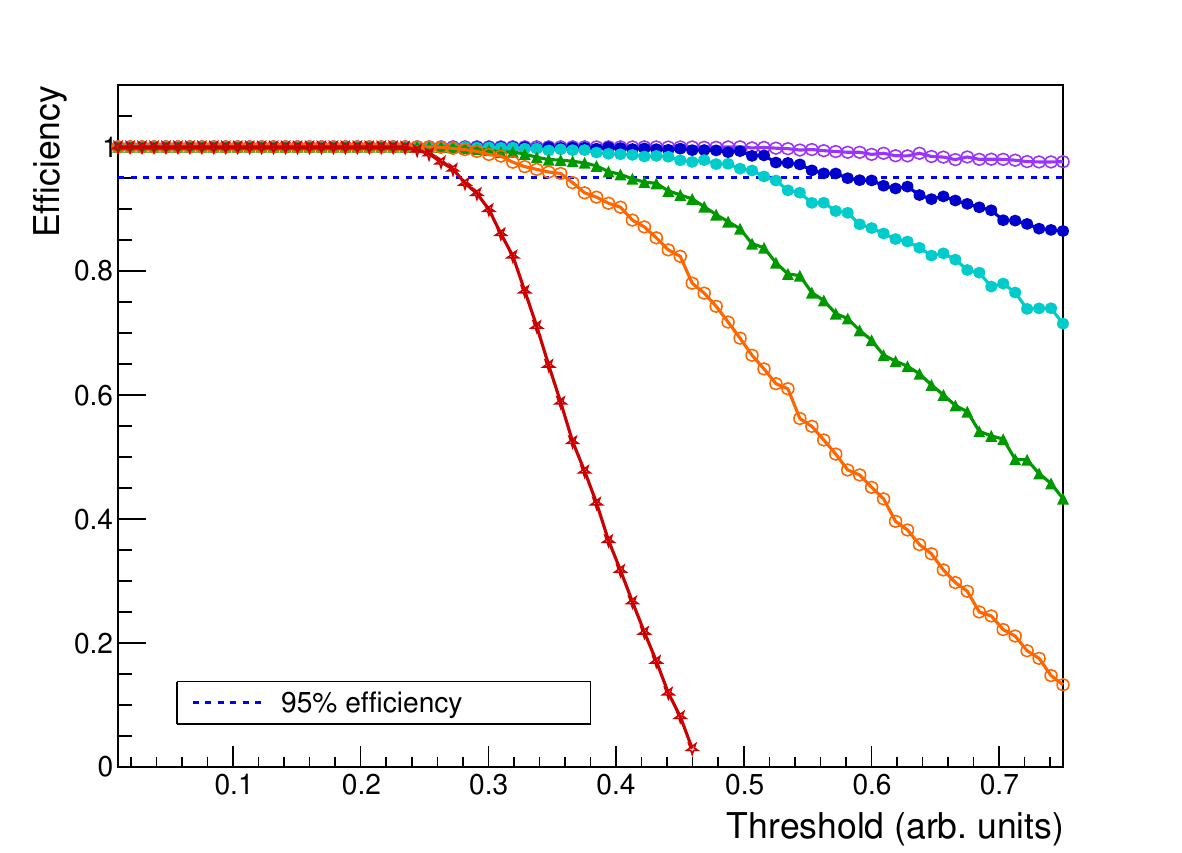}

\caption{Performance of the two-dimensional toy model. Top left: spatial resolution as a function of threshold. Top right: spatial resolution as a function of the average cluster size together with the phenomenological parameterisation from Eq.~\ref{Eq.:PhenoParameterisation}. Bottom left: average cluster size versus threshold. Bottom right: detection efficiency versus threshold. All quantities except the efficiency are shown only in the region where the detection efficiency exceeds $95\%$.}
\label{fig:2D_performance}
\end{figure}

As in the one-dimensional case, systematic scans of the threshold allow the study of spatial resolution, cluster size, and detection efficiency. The corresponding results for the two-dimensional model are summarized in Fig.~\ref{fig:2D_performance}. The figure shows the spatial resolution as a function of the threshold, the resolution as a function of the average cluster size, the average cluster size as a function of threshold, and the detection efficiency versus threshold. To ensure that the reconstructed performance is evaluated in a regime of high detection efficiency, all quantities except the efficiency itself are shown only for threshold values corresponding to efficiencies greater than $95\%$. Consistent results are obtained when $(x_{\mathrm{true}}, y_{\mathrm{true}})$ is generated uniformly over the full $21 \times 21$ pixel matrix, demonstrating that edge effects are negligible for the observables considered.

Fig.~\ref{fig:2D_performance} demonstrates that the average cluster size is an excellent scaling variable for resolution studies: independent of charge cloud width, the same spatial resolution is obtained for a given average cluster size. Motivated by this observation, we fit the resolution as a function of average cluster size $\langle N_\text{pixels} \rangle$ with a purely phenomenological 5th order polynom, intended as a practical tool for detector simulation and design studies:

\begin{equation}
\begin{split}
    \sigma_p = p \cdot \Bigl( & \frac{1}{\sqrt{12}} + a_1(\langle N_\text{pixels} \rangle-1) + a_2(\langle N_\text{pixels} \rangle^2-1) \\
    & + a_3(\langle N_\text{pixels} \rangle^3-1) + a_4(\langle N_\text{pixels} \rangle^4-1) \\
    & + a_5(\langle N_\text{pixels} \rangle^5-1) \Bigr) \; , 
     1\leq \langle N_\text{pixels} \rangle \leq 4.5
\end{split}
    \label{Eq.:PhenoParameterisation}
\end{equation}

\noindent with parameters $a_1 = -0.05848$, $a_2 = -0.2600$, $a_3 = 0.1579$, $a_4 =  -0.03192$, and $a_5 = 0.002120$.

\section{Comparison to experimental data}

\begin{figure}[h]
 \begin{center}

   \includegraphics[width=\linewidth]{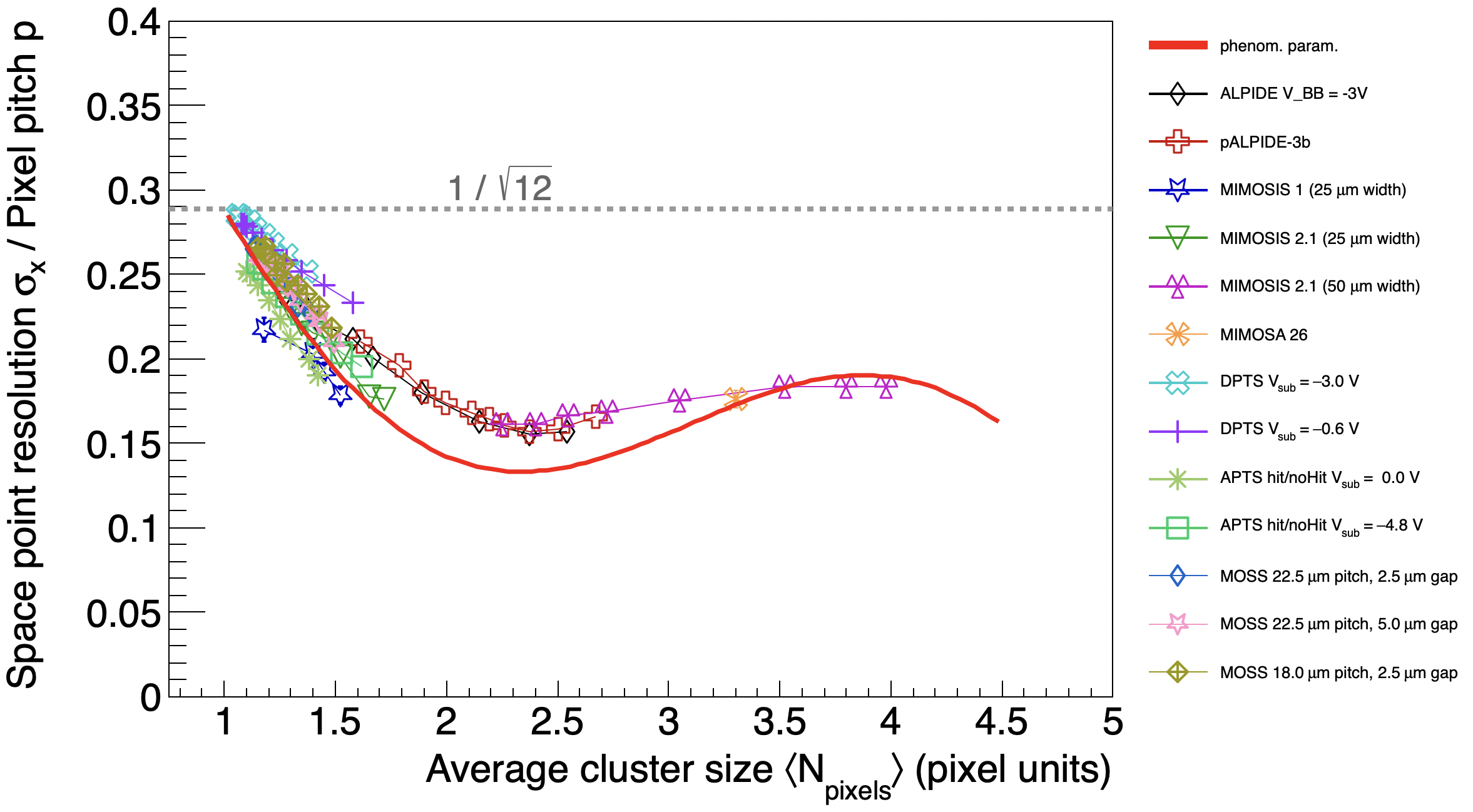}

  \caption{Comparison between data and simulation (phenomenological parameterisation from Eq.~\ref{Eq.:PhenoParameterisation}) of the spatial resolution versus the average cluster size. Data points are taken from various sources (ALPIDE~\cite{ALICE:2023udb}, pALICE-3b~\cite{Suljic:2016bmm}, MIMOSIS~\cite{Deveaux2025}, MIMOSA~\cite{Jansen:2016bkd}, DPTS~\cite{Rinella:2022htp}, APTS~\cite{Rinella:2024jzi}, MOSS~\cite{Abdelrahman:2025vkk}). }
  \label{Fig.:DataModelComparison}
 \end{center}
\end{figure}

The resolution as a function of average cluster size derived above can be directly confronted with experimental data. In many testbeam campaigns, both parameters are simultaneously measured for detectors such as ALPIDE~\cite{ALICE:2023udb}, pALPIDE-3b~\cite{Suljic:2016bmm}, MIMOSIS~\cite{Deveaux2025}, MIMOSA~\cite{Jansen:2016bkd}, DPTS~\cite{Rinella:2022htp}, APTS~\cite{Rinella:2024jzi}, and MOSS~\cite{Abdelrahman:2025vkk}. Figure~\ref{Fig.:DataModelComparison} shows the phenomenolog ical parameterisation of Eq.~\ref{Eq.:PhenoParameterisation} together with recent data from various testbeams. Naturally, this comparison is limited to detectors with (approximately) square pixels and is subject to several simplifications. In case of small differences ($<$~20\%) between the pixel pitch in X and Y direction (as for instance for the ALPIDE~\cite{Mager:2016yvj}), the arithmetic average of the values was used. Systematic and statistical uncertainties are in some cases either not reported or not available in machine-readable format; they are typically of the order of 10~\%. Although most experimental studies use the same cluster-finding algorithm as ours (i.e. only pixels that share an edge are grouped together, not diagonally adjacent ones), the clustering algorithm is not always reported, which may introduce further discrepancies. In addition, some of the data shown are still preliminary.

Despite these limitations, reasonable agreement between data and simulation is found as shown in Fig.~\ref{Fig.:DataModelComparison}. We thus conclude that the phenomenological parameterisation can serve as a useful tool for practitioners, in particular for detector simulations as well as design optimization studies. At the same time, we encourage experimenters to report their testbeam data in the format shown in Fig.~\ref{Fig.:DataModelComparison}, as this representation provides a detector-independent benchmark.

\section{Summary and outlook}

We have presented a simple -- arguably the simplest possible -- toy model to describe the space-point resolution of silicon pixel detectors with digital readout. Starting from analytical calculations for one-dimensional geometries, we demonstrated that the optimal resolution achievable through charge sharing with binary readout is $\sigma_p = \frac{1}{2\sqrt{12}}p$, corresponding to half the single-pixel limit. This result is independent of the cluster size and arises whenever the average cluster size takes a half-integer value.

A two-dimensional toy Monte Carlo simulation confirmed that the average cluster size serves as an excellent scaling variable for the resolution, regardless of the charge-cloud width. Based on this observation, we derived a practical phenomenological parameterization that was validated against testbeam data from several detector technologies, showing good agreement.

It must be clearly noted that our simplified model ignores several features present in real-life detectors. These include in particular: pixel-by-pixel threshold variations, fluctuations in the energy loss as well as in the detection process, and electronic noise. It is, however, reasonable to expect that all these features will lead to a worsening of the resolution and thus our findings represent the theoretically achievable limit for a centre-of-gravity method applied to an analogue readout detector. 

In a preliminary study, we modeled pixel-to-pixel threshold variations by smearing the threshold of each pixel with a Gaussian distribution whose standard deviation was 5\%, 10\%, or 15\% of the nominal threshold. This leads to a slight degradation of the scaling with cluster size, with deviations of up to about 7\%, but the overall scaling remains intact.

The influence of electronic noise is likely negligible for practical purposes. Modern detectors must suppress noise strongly to limit data volumes and to avoid fake clusters being picked up by the tracking algorithm. Consequently, the probability that a noisy pixel is randomly attached to a true cluster, which would artificially increase its size, should be very small. Nevertheless, quantitative studies are warranted in future work.
 
It is therefore planned to extend the presented model to fully incorporate some of these missing features. In addition, other pixel geometries -- including rectangular and hexagonal shapes -- as well as non-perpendicular track incidence angles will be investigated.

\section*{Acknowledgements}
The authors would like to acknowledge stimulating discussions with Lukas Lautner in the initial phase of the project and with the EP-ALI-DD group at CERN. The research of MM was founded by the QUASIMODO research grant (Quantum sensing for one health, CUP: H97G23000100001).

\section*{Data availability}
The data shown in this manuscript is available on reasonable request.

\bibliographystyle{unsrt}
\bibliography{references}

\end{document}